# Optically Modulated Propulsion of Electric Field Powered Photoconducting Janus Particles

Matan Zehavi[1], Daniel Sofer[1], Touvia Miloh[2], Orlin Velev[3] and Gilad Yossifon[1,2*]

[1]*Faculty of Mechanical Engineering, Micro- and Nanofluidics Laboratory, Technion–Israel Institute of Technology, Technion City 32000, Israel*

[2]*School of Mechanical Engineering Tel Aviv University Ramat Aviv 69978, Israel*

[3]*Department of Chemical and Biomolecular Engineering NC State University Raleigh, NC 27695, USA*

Herein we demonstrate the ability to optically tune the mobility of electrically powered Janus particles (JP) that are half coated with various Zinc Oxide (ZnO) semiconducting layers, i.e. polycrystalline, amorphous and amorphous with a $SiO_2$ passivation layer. The ZnO semiconductor photo-response enables increase in its electrical conductivity with light having wavelengths of sufficient photon energy with respect to the semiconductor bandgap. This effect, termed optically modulated electrokinetic propulsion (OMEP), can be harnessed to increase the contrast in polarizability between the dielectric and semiconducting hemispheres, which in turn, results in an increased electrokinetic mobility. The addition of optical activation to the electrical field enables an additional degree of control of JP mobility. We also demonstrate optical control of collective behavior and particle-particle interactions for dense semi-conducting Janus particle populations.

Active particles have emerged as a topic of scientific interest due to their ability to convert energy from their environment into autonomous translational and/or rotational motion ("self-propulsion") [1]–[3]. The underlying mechanism driving such motion is encoded in their design, as they asymmetrically draw and dissipate energy, thus creating local field gradients resulting in particle mobility [4]. For non-active particles, gradients of externally imposed macroscopic fields could lead to particle motion by effects such as electrophoresis [5], dielectrophoresis [6], magnetophoresis[7] and thermophoresis [8], [9]. However, these all result in a distinct phoretic motion, i.e., migrating en masse in an externally prescribed direction along the applied field gradients, in clear contrast to self-propelling particles, where each particle is free to travel along its individual trajectory. Active particles are a subject of large research interest and focus, due to their promise in diverse applications, such as remote microsurgery, medical analytics, self-repairing systems, and self-motile devices [10], [11].

Electric fields are a facile and controllable source of energy for powering the motion of active particles as these fields can be modulated in both frequency and intensity. This enables the precise tuning of the induced propulsion forces exerted on particles in real time (affecting particle motion) and can also simultaneously provide means for controlling particle-particle interactions [12]. Changes in the frequency of the applied electric fields can switch between several distinct electrokinetic propulsion mechanisms, such as electrohydrodynamic (EHD) flows [13], induced- charge electrophoresis (ICEP) [14], self-dielectrophoresis (sDEP) [15] and self-electrophoresis by diode rectification[16]–[18]. By rationally designing particles that respond to ambient AC electric fields using these mechanisms at different frequencies, it is thus possible to make them to change multimodally their direction of motion on demand [19].

To supplement the electrical control parameters of frequency and amplitude, we introduce here an additional means of control. This is achieved by using Janus microspheres that are half coated with semiconductor coatings and illuminated by light of varying intensities and wavelengths. Semiconducting materials are photoconductive, implying that their electronic conductivity is generally increased upon absorption of photons with sufficient photon energy. In the context of electrokinetic propulsion, the photoconductive effect allows us to change and control the amount of free charge carriers within the semiconducting layer, thereby increasing its electrical conductivity and ideally altering its electrical response from that of dielectric to conductor. The change in the electrical polarizability under an externally applied electric field, results in a change of particle mobility through the above-mentioned electrokinetic propulsion mechanisms, a phenomenon we term optically modulated electrokinetic propulsion (OMEP). Previous studies involving optical control of photosensitive semiconducting materials under an applied electric field include electro-rotation (ROT) studies of silicon nanowires [20], [21]. Studies have also been conducted where particle collective behavior can affect their optical response [22]. Optical activation of active particles adds another degree of control that can be used for addressing only a sub-population of active particles within a larger heterogeneous population.

The fabrication of the semiconducting Janus particles is based on modification of Silica microspheres of 5 µm in diameter (Polysciences Inc., Cat. No. 24332) that were deposited over a standard microscope slide. Following a 1 min $O_2$ plasma cleaning, the Janus metal/semiconductor coating was deposited. ZnO coatings are 100 nm thick sputter-deposited from a 99.95% pure ZnO target, chamber pressure of 3 mTorr and 100 W gun power. The polycrystalline ZnO was deposited at T=300 ˚C on a rotating stage, while the amorphous ZnO was deposited in room temperature in a direct vertical deposition. On some of the amorphous samples, an additional 10 nm thick layer of SiO$_2$ was deposited (room temperature deposition, 200 W gun power, chamber pressure 3 mTorr). For the metallic coating, 10 nm of Ti and 35 nm of Au was deposited using an E-beam evaporator. Following coating deposition, the particles were released and suspended for at least 2 hours in a low-concentration KCl solution 3 µS/cm, pH=6.4, with 0.05% vol/vol. Tween20 surfactant to reduce particle adhesion to the floor of the test chamber [23].

The test chamber is comprised of two glass slides separated by a 360 µm thick adhesive tape. The bottom slide is a 1.1 mm thick microscope slide covered by 4-10 Ω/sq. ITO layer (Delta technologies, Cat. No. CB-40IN-0111) with a 15 nm sputtered SiO$_2$ layer on top for suppression of particle adsorption. The top slide is a 0.7 mm thick microscope slide covered by 4-10 Ω/sq. ITO layer (Delta technologies, Cat. No. CB-40IN-1107) with drilled holes of 1 mm in diameter for inlet and outlet (see supporting material Fig. S1). Copper conductive tapes were used to connect the ITO coated glass slides to the function generator. The test chamber was placed onto a Nikon TI inverted microscope with an ANDOR iXon3 CMOS camera and a Yokogawa CSU-X1 spinning disk confocal system. The image light was introduced using the microscope's bright field light source filtered through a green interference (GIF) filter. The UV light source is a Prizmatix Mic-LED-365L, with peak wavelengths of 365 nm, introduced into the light path through a 400 nm dichroic mirror. Electric field was introduced using an Agilent 33250A function waveform generator amplified with a Falco systems WMA300 amplifier and filtered through a 10 µF capacitor. Signal parameters were monitored for integrity using a Tektronix TPS 2024 oscilloscope (see supporting material Fig. S2). For light power density measurements, the sample was replaced with a photodiode (Ophir photonics PD300 UV sensor) through a 150µm diameter circular aperture, measured with an Ophir photonics VEGA power meter. Light power measurements for the mobility experiments were performed through the same ITO-coated slide used in the test chamber (Delta technologies, Cat. No. CB-40IN-1107).

The electrokinetic mobility of a Janus particle relies on the contrast between the electrical polarizabilities of its opposite hemispheres. The electrical conductivity of the ZnO coated JP side is changed due to illumination by light of wavelengths that have sufficient photon energy relative to the coating band gap. The active particle mobility can be thus controlled by means of ICEP (Fig. 1a) and sDEP electrokinetic propulsion mechanisms. We coated the Janus particles with polycrystalline ZnO (Fig. 1b), a semiconductor material with a band gap of ~3.4 eV [24] to achieve response to light activation with wavelengths under 365 nm. Fig. 1c shows a particle's velocity response to UV under fixed electric field amplitude and frequency, when suspended in low conductivity (3 µS/cm) KCl solution. As clearly seen the particle velocity increases significantly with UV illumination. The small velocity with no UV is a result of the polycrystalline non-negligible ZnO dark conductivity (see Fig.2c). No mobility response was observed when applying UV light with no external electric field (see Video S1). Therefore, self-thermophoretic [25], [26] and photocatalytic [27] mechanisms can be excluded as a cause for particle propulsion. To illustrate the transition between different electrokinetic propulsion modes, Fig. 1d shows a Janus particle path traced under frequency switching between 5 kHz and 100 kHz subject to fixed UV illumination and electric field amplitude. The propulsion switching mechanisms corresponding to ICEO and sDEP act in opposite directions.

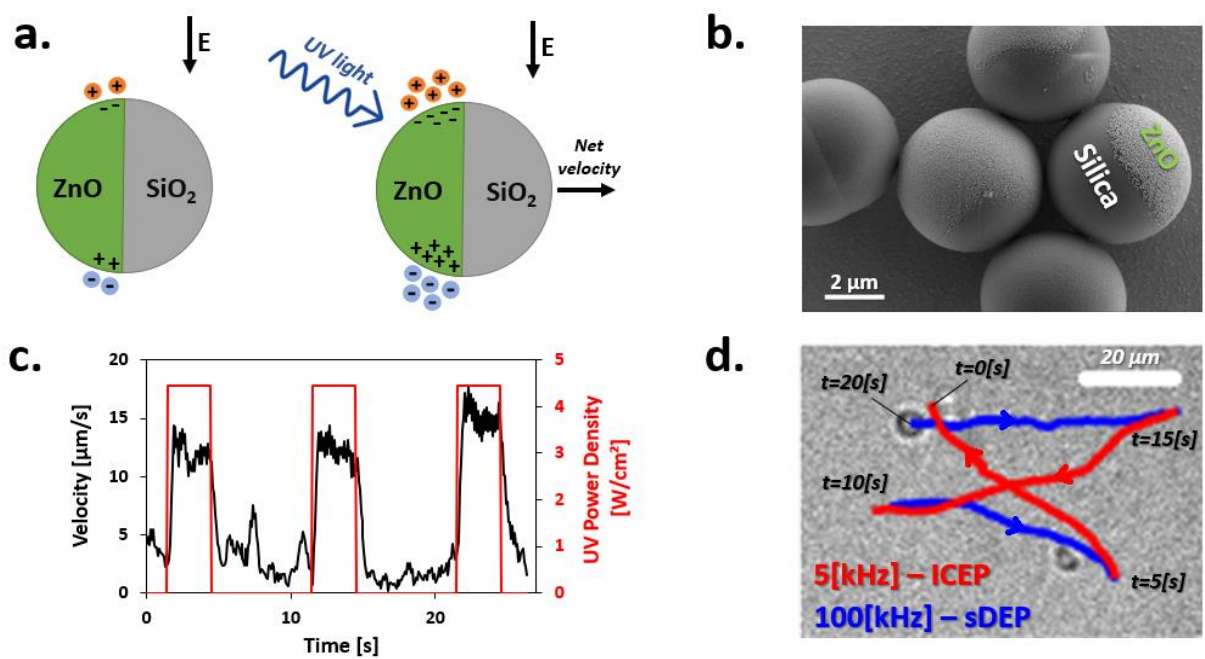

**Fig. 1:** Optically modulated electrokinetic propulsion (OMEP). (a) Schematics of the optical (UV) gated mobility of electrically powered semi-conducting Janus particles at low frequencies corresponding to ICEP mode; (b) SEM image of silica microparticles half coated with a 100 nm thick layer of polycrystalline ZnO; (c) A representative velocity plot of a Janus particle at 8 kHz, 166.7 kVpp/m, electric field under UV (365 nm) light switching; (d) JP trajectory under continuous illumination of UV light with electric field frequency switching between 5 kHz and 100 kHz, corresponding to ICEO and sDEP propulsion modes, respectively, acting in opposite directions. See supplementary video S1 for representative mobility videos.

To directly measure the ZnO layer conductivity response to light, a photoresistor test structure was fabricated (inset of Fig. 2a) using common photolithography techniques (see supporting Fig. S3 for manufacturing schematics). The photoconductive layer was deposited onto a 1 mm thick glass slide and patterned into a 0.5×0.5 mm$^2$ photoresistor, in the same deposition process as the Janus particles. Electrical contacts of 50 nm thick Ti followed by 200 nm thick Au were deposited by an ion-beam evaporator and soldered to wires for connectivity. The coupon was exposed to different intensities of UV light and its electrical behavior was evaluated using I-V measurements, using a Kieithley 2636A Source Meter (see supporting material Fig. S4). To examine the wavelength dependency (quantum effect) of the photoconductive response, we illuminated the photoresistor with varying wavelengths at the same power density and measured its conductivity (Fig. 2b). It can be clearly seen that for wavelengths higher than 365 nm, which correspond to the ZnO bandgap energy, the photoconductivity decreases by few orders of magnitude, as can be expected from ZnO absorption [28]. However, for wavelengths near the bandgap the photoconductivity response was found to be sufficient for the JP to clearly exhibit propulsion (see supporting materials Fig. S5 for the mobility response of a polycrystalline ZnO JP under an illumination of 390 nm).

As clearly seen in Fig. 2c the electronic conductivity of the semiconductor layer can be continuously tuned via variation of the power density of the UV illumination. We have further demonstrated the ability to tune the mobility of a Janus particle via optical gating of varying optical power densities under a continuous fixed electric field (Fig. 2d, Supplementary Video S1).

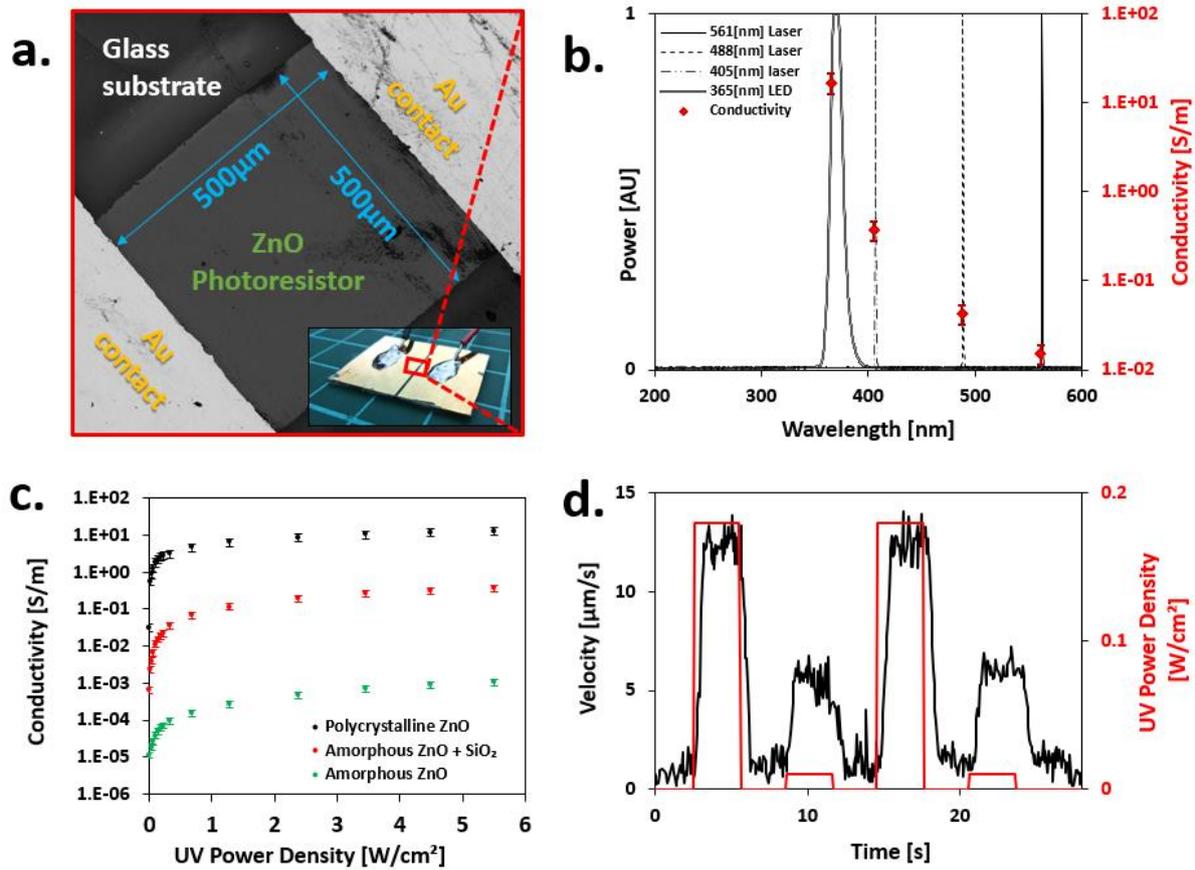

**Fig. 2:** Optical control over the electrical conductivity and Janus particle mobility. (a) SEM image of the conductivity test structure used in the photoconductivity measurements. inset: test chip photo; (b) Conductivity measurements of polycrystalline ZnO test structure under various illumination sources at 0.14 W/cm² and their source spectrum; (c) Electrical conductivity measurements of different types of ZnO photoresponsive coatings under varying UV (365 nm) illumination intensity. The measurement corresponding to zero UV power density was taken under ambient light conditions; (d) Effect of dynamic optical gating of a polycrystalline ZnO JP mobility at varying optical power densities, under constant application of an electric field of 100 kHz and 166.7 kVpp/m. The error bars in Fig.2b,c represent the standard deviation of ZnO layer thickness that is inversely related to the measured conductance.

Fig. 3a depicts the frequency dispersion of the JP mobility under a constant applied AC electric field of 166.7 kVpp/m and UV activation power density of 4.4 W/cm². The photoconductive particles show a similar frequency response behavior to that of the controlled metallo-dielectric ('Au') particles, demonstrating distinct low-frequency ICEP and high-frequency sDEP regions. In correspondence to the photoconductive responses measured for the different coating types, the mobility of the ZnO Janus particles varies with the coating conductivity. The data in Fig. 3b show increasing JP mobility by optically changing the semiconductor conductivity. These experimental results are in qualitative agreement with the theoretical predictions in [29] for ICEP. For the sDEP region, results stand in qualitative agreement with numerically calculated JP velocity (see inset of Fig.3b), exhibiting a monotonic increase in magnitude with increasing electrical conductivity $\sigma_i$ of the semi-conducting coating (see simulation details in the SI).

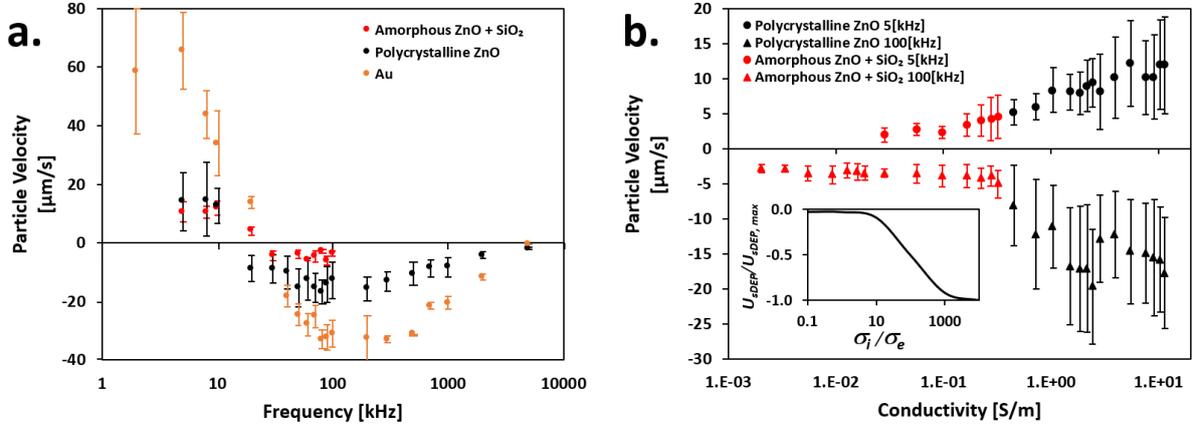

**Fig. 3:** Frequency dispersion of the JP mobility and its dependency on the optical illumination intensity. (a) Mobility of JPs with different coatings versus frequency under an applied electric field of 166.7 kVpp/m and 4.4 W/cm$^2$ UV intensity (365 nm); (b) Dependency of JPs mobility on ZnO conductivity (subject to UV intensity; see Supplementary Figure S6) at applied electric field frequencies of 5 kHz and 100 kHz under 166.7 kVpp/m and corresponding to ICEP and sDEP modes, respectively. The transition from UV intensity to electrical conductivity was based on the experimental measurements depicted in Fig.2c. Inset: Numerically calculated normalized JP velocity (under sDEP mode at a normalized frequency of $\widetilde{\Omega} = \frac{\omega \lambda_0 a}{D} = 100$) for varying ratio of semi-conducting coating ($\sigma_i$) to electrolyte solution ($\sigma_e$) conductivities. Error-bars are velocity standard deviation.

The collective behavior of a large-density population of amorphous ZnO JPs with SiO$_2$ passivation as well as polycrystalline ZnO JPs and bare silica beads was explored by applying a fixed-amplitude electric field with low (8kHz) and high (1MHz) applied frequencies while transitioning between UV on/off states. The images in Fig. 4 depict the centers of the three different particle types as well as the path that they moved along within a 5 second period to give a qualitative sense of their mobility and collective behavior (see the corresponding MSD (mean square displacement) analysis in Fig.S7). To avoid the transients upon application of an electric field with UV off, analysis begins 5 seconds after the application of the field. At the low-frequency regime and upon electric field activation with UV off, the JPs behave as dielectric silica spheres and self-assemble into a structure with uniform inter-particle spacing. Upon UV illumination, the JPs still exhibit a repulsive dipolar interaction with neighboring particles (due to the induced EDL screening of the photoactivated semiconducting hemisphere), while some of the JPs move due to ICEP effects.

At the other extreme end of the high- frequency region, as demonstrated in the 1 MHz case, there is practically no induced EDL screening of the JPs. Therefore, the JP can be considered as an entity including two dipoles of opposite directions, corresponding to the dielectric and semiconducting sides. With no UV activation, the JP's collective response is similar to that of the 8 kHz case as the semiconducting hemisphere is not photoactivated, and the dielectric dipole dominates the dipolar repulsion. However, under UV exposure we observe an attractive dipolar interaction between the dipole of the dielectric JP side with the corresponding dipole of its neighbor's photoactivated semiconducting side, resulting in a chain formation. These are in a qualitative agreement with previous findings [29]-[31] which also observed such a collective behavior but with metallic (Au) coated JPs at the low and high frequency extremes. They suggested that it is the dielectric-dependent frequency mismatch between the two JP hemispheres that allows simultaneous control of particle mobility and colloidal interactions. For a similar ITO parallel plate system, [30] showed that it is possible to reconfigure active particles into various collective states by introducing imbalanced interactions to realize swarms, chains, clusters and isotropic gases from the same precursor particle by changing the electric-field frequency.

To further understand the effect of optical modulation on the JPs collective behavior, we generalized Yan et al. [30] JP-JP dipolar interaction model to account for the electric field frequency and complex permittivity of the semi-conducting hemisphere with varying electrical conductivity (see supporting materials). The model is in qualitative agreement with the experiments in the sense that it succeeds to predict the repulsive interaction at low frequencies (regardless of UV illumination) as well as the transition to attractive interaction (i.e. formation of chains) with increasing frequency and UV illumination (see Fig.S9 in supporting materials).

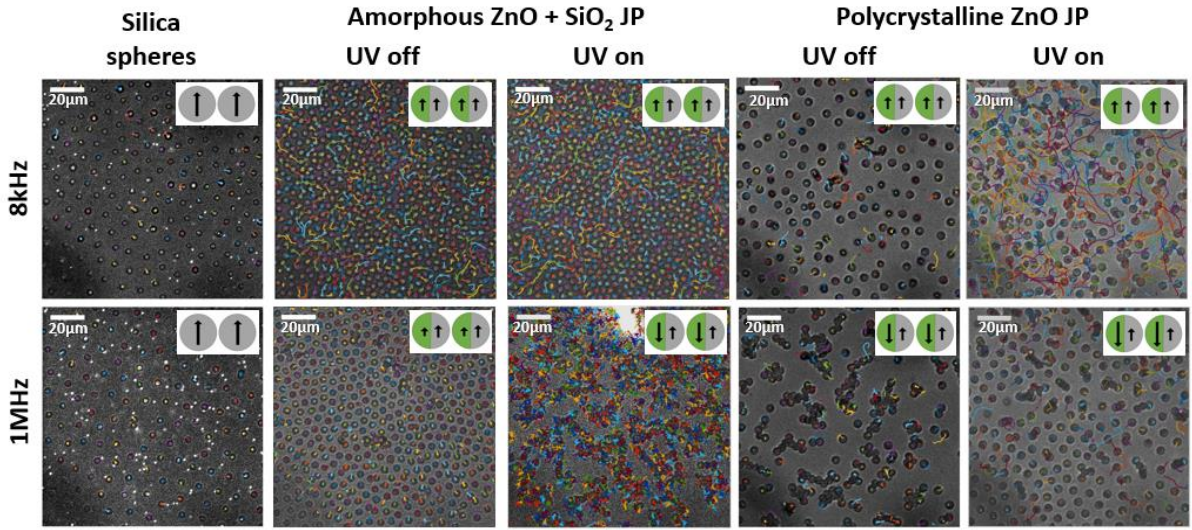

**Fig. 4:** Collective behavior of highly dense JPs of different types shown for 8kHz and 1MHz frequencies under 166.7 kVpp/m electric field, with/without UV activation (7 W/cm²). Path lines drawn for a period of 5 s. Corresponding MSD (mean square displacement) plots that provide a qualitative measure of the JPs collective motion are depicted in the supporting information. See Supplementary Figure S7 and videos S2, S3.

To summarize, we have introduced a novel approach for controlling the mobility of electrokinetically driven active Janus particles and modulating their collective behavior, by means of optically controlling the electrical properties of their semi-conducting half-sphere coating. Specifically, we used silica particles that were half-coated with photoconductive ZnO layers to form a Janus particle with semi-conducting side. We have characterized the photoconductivity effect using a test chip for several coating types (polycrystalline ZnO, amorphous ZnO, and amorphous ZnO with a passivation $SiO_2$ layer) for varying optical illumination intensities and wavelengths. We then proceeded to measure the velocity of electrokinetically driven JPs under varying electric fields and UV power densities. The results showed that the added degree of UV light control enables not just an 'on-off' switching but also continuous tuning of the electrokinetic ICEP and sDEP velocities for isolated particles. The optically modulated electrokinetic propulsion also allows efficient controlling and switching between different collective behavior types in dense particle populations, by tuning the JP particle induced dipoles and the resulting dipolar interaction between interacting JPs. This control of OMEP enables new degrees of freedom in active semi-conducting particle design as they can be selectively activated within a heterogenous population of particles by judicious selection of the excitation wavelength (i.e., according to the bandgap of the different semi-conducting coatings). Combining different semiconducting coatings onto the same active particle can also enable to control not only its velocity but also its trajectory as well as the collective interactions of systems of such engineered active particles.

M.Z, G.Y., T.M and O.V wish to acknowledge support from the Binational Science Foundation (BSF) Grant 2018168. O.V. also acknowledges partial support from NSF CBET-2133983. We wish to acknowledge the Technion Russel-Berrie Nanotechnology Institute (RBNI) and the Technion Micro-Nano Fabrication Unit (MNFU) for their technical support. We thank Prof. Yosi Shacham-Diamand from Tel-Aviv University for his valuable advice and inputs regarding the photoconductive effects and processes.
* Corresponding author. gilad.yossifon@gmail.com